\documentstyle[prl,aps,multicol]{revtex}
\renewcommand{\narrowtext}{\begin{multicols}{2}
\global\columnwidth20.5pc}
\renewcommand{\widetext}{\end{multicols} \global\columnwidth42.5pc}

\begin{document}
\draft
\title{Energy level dynamics in systems with weakly multifractal
eigenstates: equivalence to  1D correlated fermions at low
temperatures}
\author{V.E.Kravtsov}
\address{The Abdus Salam International Centre for Theoretical Physics,
P.O.B.
586, 34100 Trieste, Italy, \\ Landau Institute for Theoretical
Physics, 2 Kosygina st., 117940 Moscow, Russia}
\author{A.M.Tsvelik}
\address{Department of Physics, University of Oxford
1 Keble Road, Oxford, OX1 3NP, United Kingdom}
\maketitle
\begin{abstract}
It is shown that the parametric spectral statistics in the critical random
matrix
ensemble with multifractal eigenvector statistics are identical to the
statistics of correlated 1D fermions at finite temperatures. 
For weak
multifractality the effective  temperature of fictitious 1D
fermions is proportional to $T_{eff}\propto (1-d_{n})/n \ll 1$, where
$d_{n}$ is the
fractal
dimension found from the $n$-th moment of inverse participation ratio.
For large energy and parameter separations the fictitious fermions are
described by
the  Luttinger liquid model which follows from the Calogero-Sutherland
model. The low-temperature asymptotic form of the two-point
equal-parameter spectral 
correlation
function is found for all energy separations and its relevance for the low
temperature equal-time density correlations in the Calogero-Sutherland
model is conjectured.   

\end{abstract}

\narrowtext

The spectral statistics in complex quantum systems are signatures of the
underlying dynamics of the corresponding classical counterpart. The
spectral statistics in chaotic and disordered systems in the limit of
infinite dimensionless
conductance $g$ is described by the
classical random matrix theory of Wigner and Dyson
\cite{RMT} (WD statistics).  The WD statistics possess a remarkable
property of universality: it depends only on the symmetry class with
respect to the time-reversal transformation ${\cal T}$. The three symmetry
classes correspond to the lack of ${\cal T}$-invariance (the unitary
ensemble, $\beta=2$); the ${\cal T}$-invariant systems with ${\cal T}^2
=1$ (the orthogonal ensemble, $\beta=1$), and  the ${\cal T}$-invariant
systems with ${\cal T}^2   
=-1$ (the symplectic ensemble, $\beta=4$), respectively. The physical
ground behind this universality is the structureless eigenfunctions in the
ergodic regime which implies the invariance of the eigenfunction
statistics with respect to a unitary transformation of the basis.

In real disordered metals the eigenfunctions are not basis-invariant. The
basis-preference reaches its extreme form for the strongly impure metals
where all eigenfunctions are localized in the coordinate space but
delocalized in the momentum space. In this case the spectral statistics is
Poissonian in the thermodynamic (TD) limit.

For low-dimensional systems $d=1,2$, where all states are
localized in the TD limit, one can observe the smooth crossover
from the WD to the Poisson spectral statistics as a function of the
parameter $\xi/L$, where $\xi$ is the localization radius and $L$ is the
system size. The dependence of the spectral correlation functions on the
energy variable $s=E/\Delta$ ($\Delta$ is the mean level separation) 
is non-universal for
finite
$L/\xi$ but all of them tend to the Poisson
limit
as $L/\xi\rightarrow\infty$.

In systems of higher dimensionality $d>2$ the situation is different
because of the presence of the Anderson localization transition at a
critical disorder $W=W_{c}$. In the metal phase $W<W_{c}$ the dimensional
conductance $g(L)\rightarrow \infty$ as $L\rightarrow\infty$
and one obtains the WD spectral statistics in the TD limit. In the
insulator state $g(L)\rightarrow 0$ at $L\rightarrow\infty$ and the
limiting statistics is Poissonian. However, there is a fixed point
$W=W_{c}$ in which the spectral statistics are nearly independent of $L$.
Thus at the critical point there exist universal spectral
statistics which are neither WD
nor Poissonian but rather a  hybrid of 
both \cite{Shkl}. However, the universality of the critical spectral
statistics (CSS) is somewhat limited, since it depends not only on the
Dyson symmetry parameter $\beta$ but also on the critical value of the
dimensional conductance $g^{*}$ which in turn depends on the 
dimensionality $d$ of the system \cite{Zhar1}. Thus for each
universality class there is a {\it
family} of critical spectral statistics parametrized by the critical
dimensionless conductance $g^{*}$.

The very existence of the subject of critical level statistics imposes a
constraint on the possible values of the localization length exponent
$\nu=[d\beta(g)/d\ln g]^{-1}|_{g=g^{*}}$, where $\beta(g)\equiv d\ln
g/d\ln L$ is the scaling 
function. Indeed, for the spectral statistics to be meaningful
the width
of
the critical energy window $\delta E$  must be much larger than the mean
level separation $\Delta \propto 1/L^{d}$. The quantity $\delta E$ is
defined as the
distance from the mobility edge $E=E_{c}$ at which the
localization or correlation radius
$\xi(\delta E)\propto |\delta E|^{-\nu}$ is equal to the system size $L$.
The number of critical eigenstates ${\cal
N}=\delta E/\Delta$ is proportional to $L^{d-\frac{1}{\nu}}$. 
For 
$\nu d >1$
this number tends to infinity in the
TD limit $L\rightarrow\infty$ despite the width of the critical energy
window shrinks to zero. This necessary condition for the existence of the
critical statistics is secured by the famous Harris criterion $\nu d
> 2$.

However, the critical exponent $\nu$ enters not only in the necessary
condition for the CSS but also in the correlation functions of the
density of energy levels $\rho(E)$.
It has been shown in \cite{KLAA,AKL} that there is a power-law tail in the
critical 
two-level correlation function (TLCF)
$R(\omega)=\langle\langle\rho(E)\rho(E+\omega)
\rangle\rangle$ that arises because of the
finite-size
correction to the dimensionless conductance
$g(L_{\omega})/g^{*}-1\propto
(L_{\omega}/L)^{1/\nu}= s ^{-\frac{1}{\nu d}}$, where
$L_{\omega}\propto \omega^{-1/d}\ll L$ is
the length scale set
by the
energy difference $\omega=E-E'\equiv s \Delta$ between two levels.
The sign of this tail depends  on whether the critical energy levels are
on the metal ($E<E_{c}$) or on the insulator ($E>E_{c}$) side of the
mobility
edge, in the same way as for the 2D systems in the weak delocalization
($\beta=4$) or weak localization ($\beta=1$) regimes \cite{KL2}. 
Clearly, this power-law tail does not reflect  properties of the
critical
eigenstates but rather the behavior $\xi(|E-E_{c}|)$ of the size of the
space region
where eigenstates show the critical
space correlations. 

In order to study the relationship between the properties of the critical
eigenstates and the CSS in its pure form one should consider a system with
a continuous line of critical points where $\beta(g)=0$. This case
formally corresponds to $\nu=\infty$ and the finite-size effects are
absent. 

Another complication which makes ambiguous the definition of CSS 
is the fact that being independent of the system size, the spectral 
correlation
functions depend on
the boundary conditions \cite{Mon,PSch,KY} and topology of a system. 
Therefore we will consider the system of the torus topology where CSS
takes its `canonical' form. In particular, the TLCF decays exponentially
in this case. 

As has been already mentioned the universality of the WD statistics
is based on the ergodic, basis-invariant statistics of
eigenfunctions which one may encounter in different physical situations.
The characteristic feature of all critical quantum systems is the
multifractal statistics of the critical eigenfunctions \cite{W,CCP,AlKL}.
The simplest two-point correlations of the critical wave functions
can be obtained from the renormalization group result
\cite{W2} for $l<r<\xi<L$:
\begin{equation}
\label{Wegn}
\langle |\Psi_{E}(0)|^{2n}|\Psi_{E}(r)|^{2n}\rangle
=p\,
<|\Psi_{E}(0)|^{2n}>^2 \,(\xi/r)^{\alpha_{n}},
\end{equation}
where $l$  is the short-distance cut-off of the order of the elastic
scattering length, $p=(\xi/L)^{d}$ is the probability for a reference
point to be inside a localization region. The exponent
$\alpha_{n}=2n(d_{n}-d_{2n})+d_{2n} -2d_{n}+d$
is  expressed through the fractal dimensions
$d_{n}$ defined by the $L$-dependence of the moments of inverse
participation ratio:
\begin{equation}
\label{mf}  
L^{d} \langle|\Psi_{E}(r)|^{2n} \rangle \propto
L^{-d_{n}\,(n-1)},\;\;\;\;\;n\geq 2.
\end{equation}
At the critical point $p=1$ and the correlation radius
$\xi\rightarrow\infty$ in
Eq.(\ref{Wegn}) must be
replaced by the sample size $L$. Spectral statistics are
related with eigenfunction correlations at 
different energies $E$ and $E'=E+\omega$. If $\omega\gg \Delta$ and
thus $L_{\omega}\ll L$ one should substitute $L_{\omega}$ for $L$ in the
$r$-dependent term of Eq.(\ref{Wegn}). In this way the multifractality
exponents $d_{n}$ enter the spectral $\omega$-dependences
\cite{Ch}.   

For weak
multifractality one can expect the fractal dimensions to be a linear
function of $n$ which is controlled by only one
parameter $a$:
\begin{equation}
\label{weak}
d_{n}/d=1-a n,\;\;\;\;\;\;\;\;a\sim 1/g^{*}\ll 1.
\end{equation}
This relationship holds approximately for the Anderson transition in
$2+\epsilon$
dimensions \cite{W,CCP,AlKL} and is fulfilled exactly for the critical
eigenstate of the Dirac equation in the random vector-potential
\cite{Mud}. 

Thus for  critical quantum systems with weak multifractality it is
natural to
expect that the spectral statistics  depends 
 on only one system-specific parameter - the critical
conductance $g^{*}$. 

In view of the expected universality, it is useful to find a simple
one-parameter random matrix ensemble with the multifractal eigenfunction
statistics which would play the same role for the critical systems as the
classical RMT does for the ergodic systems. 
As a matter of fact there are few candidates \cite{KMut}. However, here
we focus only on one of them \cite{MirFyod}, since for this ensemble 
the multifractality of
eigenstates has been rigorously proven \cite{MirFyod,AlLev}.

Consider a Hermitean $N\times N$ matrix with the real ($\beta=1$), complex
($\beta=2$)
or quaternionic ($\beta=4$) entries $H_{ij}$ $(i\geq j)$ which are
independent
Gaussian random numbers with zero mean and the variance:
\begin{equation}
\label{MFmat}
\langle|H_{ij}|^{2} \rangle =
\frac{1}{1+\frac{(i-j)^2}{B^2}}
\left\{
\begin{array}{cc}
1/\beta, & i=j, \\
1/2 , & i\neq j
\end{array}
\right.\equiv J(i-j)
\end{equation}
This model has been shown to be critical both for large \cite{MirFyod}
and for small \cite{AlLev} values of $B$ with the fractal dimensionality
$d_{2}$ at the center of the spectral band being:
\begin{equation}
\label{eta}
d_{2}=\left\{
\begin{array}{cc}
1-\frac{1}{\pi\beta B} , & B\gg 1, \\
2 B , & B\ll 1
\end{array}
\right.
\end{equation}
Thus the 1D system with long-range hopping described by the matrix
Hamiltonian Eq.(\ref{MFmat})
possesses the line of critical points $B\in (0,\infty)$, the fractal
dimensionality $d_{2}$ changing from 1 to 0 with decreasing $B$.

One can extend this matrix 1D model by closing it into a ring and applying
a `flux' $\varphi\in [0,1]$. In this case
\begin{equation}
\label{fl}
H_{ij}(\varphi)=H_{ij}+H_{ij}^{(1)}\,e^{2\pi i\varphi\, sgn(i-j)} 
\end{equation}
is a
sum of
two independent
Gaussian
random numbers with the variance of $H_{ij}^{(1)}$ given by:
\begin{equation}
\label{ring}
\langle|H_{ij}^{(1)}|^{2} \rangle = J(N-|i-j|).
\end{equation}

For large values of $B$ which correspond to weak multifractality
one can derive \cite{MirFyod} an effective field theory -- the
supersymmetric nonlinear sigma-model \cite{Ef} -- which describes the
spectral and eigenfunction correlations of the critical random matrix
ensemble Eq.(\ref{MFmat}):
\begin{equation}
\label{sigma}
F[{\bf Q}]= -\frac{ g^{*}}{16}\,\sum_{i,j=1}^{N}\,Str\,
[{\bf Q}_{i}\,U_{|i-j|}\,{\bf Q}_{j}]+\frac{i\pi
s}{4N}\,\sum_{i=1}^{N}\,Str[\sigma_{z}
{\bf Q}_{i}],
\end{equation}
where ${\bf Q}$ is the supermatrix with ${\bf Q}_{i}^{2}={\bf 1}$ and
\begin{equation}
\label{g}
g^{*}= 4 \beta B.
\end{equation}
The symmetry with respect to time reversal is encoded in the symmetry of
${\bf Q}_{i}$ in exactly the same
way as for the
diffusive sigma-model \cite{Ef}. The only difference is the long-range
kernel $U_{|i-j|}$ with the Fourier-transform $\tilde{U}_{k}=|k|$.
For a torus geometry $k=2\pi m/N$, where $m$ is an {\it arbitrary}
integer.
 
One can explicitly resolve the constraint ${\bf Q}^{2}=1$ by switching to
the
integration over the `angles' $W$. Then the Gaussian fluctuations of
`angles' recover the spectrum of `quasi-diffusion' modes: 
\begin{equation}
\label{qdif}
\varepsilon_{m}=g^{*}\,|m|,\;\;\;\;\;m=0,\pm 
1,\pm 2,...
\end{equation}
The problem of spectral statistics can be generalized to include the
dependence of spectrum on the flux $\varphi$ introduced by Eq.(\ref{fl}).
One can define \cite{SA} the
{\it parametric} two-level correlation function
$R(s,\varphi)=\langle\langle 
\rho(E,0)\rho(E+s\Delta,\varphi)\rangle\rangle$ which can be treated in the
framework of the same nonlinear sigma-model but with the 
phase-dependent  $\varepsilon_{m}$: 
\begin{equation}
\label{phase}
\varepsilon_{m}(\varphi)=g^{*}\,|m-\varphi|. 
\end{equation}
Following the work by Andreev and Altshuler Ref.\cite{AA} we introduce the
spectral determinant:
\begin{equation}
\label{sd}
D^{-1}(s,\varphi)=\prod_{m\neq 0}
\frac{\varepsilon_{m}^{2}(\varphi)+s^2}{\varepsilon_{m}^{2}(0)}
\end{equation}
Then it can be shown in the same way as in Ref.\cite{AA} that the
parametric TLCF for $s\gg 1$ and
$g^{*}\varphi\gg 1$ can be expressed
in terms of the spectral determinant as follows:
\begin{eqnarray}
\label{AAu}
R^{u}(s,\varphi)&=&-\frac{1}{4\pi^2}\frac{\partial^2
G(s,\varphi)}{\partial
s^2}+
\cos(2\pi s)\,e^{G(s,\varphi)}\\ \label{AAo}
R^{o}(s,\varphi)&=&-\frac{1}{2\pi^2}\frac{\partial^2
G(s,\varphi)}{\partial
s^2}+2
\cos(2\pi s)\,e^{2G(s,\varphi)}\\ \nonumber
R^{s}(s,\varphi)&=&-\frac{1}{8\pi^2}\frac{\partial^2
G(s,\varphi)}{\partial
s^2}+\frac{\pi}{\sqrt{8}}\,
\cos(2\pi s)\,e^{G(s,\varphi)/2}+\\ \label{AAs} &+& \frac{1}{8}\,\cos(4\pi
s)\,e^{2G(s,\varphi)},
\end{eqnarray}
where $G(s,\varphi)=G(s,\varphi+1)$ is a periodic in $\varphi$ function:
\begin{equation}
\label{G}
e^{G(s,\varphi)}=\frac{D(s,\varphi)}{2\pi^2\, (s^2 +
\varepsilon_{0}^{2}(\varphi))}.
\end{equation}
Eqs.(\ref{AAu}-\ref{AAs}) coincide with the corresponding formulae in
Ref.\cite{AA}
for the unitary, orthogonal and symplectic ensembles 
after some
misprints are corrected as in
Ref.\cite{KamMez} and $s\rightarrow 2s$ for the symplectic ensemble 
to take account of the Kramers degeneracy. The only difference is in the
form of the spectral determinant Eq.(\ref{sd}) due to the specific
spectrum $\varepsilon_{m}(\varphi)$ of the quasi-diffusion modes.

Using the functional representation Eq.(\ref{sigma}) one can also find the
leading term in the deviation from the WD statistics
at $s\ll g^{*}$ and $\varphi=0$ using the results of Ref.\cite{KrMir,MirFyod}: 
\begin{equation}
\label{small}
\delta R(s)= \frac{1}{2\pi^2 \beta}\,\left(\sum_{m\neq
0}\frac{1}{\varepsilon_{m}^{2}}\right)\,\frac{d^2}{ds^2}\,\left[s^2\,R_{WD}(s)
\right],
\end{equation}
where $R_{WD}(s)$ is the Wigner-Dyson TLCF. 

A remarkable property  of the function $G(s,\varphi)$  for
the
critical
sigma-model Eq.(\ref{sigma}) on a torus is that
it
can be decomposed into
the sum
$G(s,\varphi)=F(z)+F(\bar{z})$ of
analytic
functions $F(z)$ and $F(\bar{z})$ where $\tau=g^{*}\varphi$,
$z=\tau+is$,
$\bar{z}=\tau-is$ and
\begin{equation}
\label{F}
F(z)= -\ln[\sqrt{2}g^{*}\,\sin(\pi z/g^{*})].
\end{equation}
Eq.(\ref{F}) results from a straightforward evaluation \cite{KMut} of the
product in
Eq.(\ref{sd}).

On the other hand, it can be easily verified  that $G(s,\varphi)$
given by Eq.(\ref{F}) is proportional to the Green's function of the
free-boson field $\Phi(z)$
on
a torus in the $(1+1)$ $z$-space:
$0<\Re z<g^{*}$, $-\infty<\Im z<+\infty$:
\begin{equation}
\label{gf}
\langle\Phi(s,g^{*}\varphi)\,\Phi(0,0) \rangle_{S} -
\langle\Phi(0,0)\,\Phi(0,0) \rangle_{S} =K\, G(s,\varphi),
\end{equation}
where $\langle ... \rangle_{S}$ denotes the functional average with the
free-boson action:
\begin{equation}
\label{bos}
S[\Phi]=\frac{1}{8\pi K}\int_{0}^{g^{*}}d\tau\int_{-\infty}^{+\infty}
ds\;\left[(\partial_{s}\Phi)^2 + (\partial_{\tau}\Phi)^2 \right].
\end{equation}
Now we are in a position to make a crucial step  and suggest
that for the critical RMT described by Eq.(\ref{MFmat}), the
Andreev-Altshuler
equations
(\ref{AAu}-\ref{AAs}) are nothing but density-density correlations
in the Luttinger liquid of fictitious 1D fermions  at a
finite temperature $T=1/g^{*}$:
\begin{equation}
\label{dd}
R(s,\varphi)=\bar{n}^{-2}\,\langle
n(s,\tau)\,n(0,0)\rangle_{S}\,-1,\;\;\;\; \tau=g^{*}\varphi.
\end{equation}
Indeed, the density operator $n(s,\tau)$ ($s$- is space and $\tau\in
(0, 1/T)$ is imaginary time coordinate) for  1D interacting 
fermions with 
the
Fermi-momentum $k_{F}=\pi$  can be expressed through
the free boson field $\Phi(s,\tau)$ as follows \cite{shura}:  
\begin{eqnarray}
\label{dens}
n(s,\tau)&=&\frac{1}{2\pi}\,\partial_{s}\Phi(s,\tau)+
A_{K}\,\cos[2\pi s +\Phi(s,\tau)]+\\ \nonumber &+& B_{K}\,\cos[4\pi s +
2\Phi(s,\tau)].
\end{eqnarray}
The constants $A_{K}$ and $B_{K}$ are independent of `temperature'
$1/g^{*}$ but
depend on the interaction constant $K$. They can be uniquely determined
from the WD limit $g^{*}\rightarrow\infty$.

Using Eqs.(\ref{dd},\ref{dens}) and the well known result for the Gaussian
average
of the
exponent:
\begin{equation}
\label{Ga}
\langle e^{ip \Phi(s,\tau)}\,e^{-ip
\Phi(0,0)}\rangle=e^{K p^2\,[G(s,\tau)-G(0,0)]},
\end{equation}
one can verify that for the choice: 
\begin{equation}
\label{K}
K=\frac{2}{\beta},\;\;\;\;\;\beta=1,2,4
\end{equation}
the Andreev-Altshuler formulae Eqs.(\ref{AAu}-\ref{AAs})
are reproduced exactly for the orthogonal, unitary and symplectic
ensembles, respectively.
Now we remind on a known result that the parametric
spectral 
statistics in the WD limit $g^{*}=\infty$ is equivalent to the 
Tomonaga-Luttinger
liquid at zero temperature. It follows directly from 
Ref.\cite{SLA}
and the equivalence of the Calogero-Sutherland model and the Tomonaga-Luttinger
liquid for large distances $|z|\gg 1$. The critical random matrix ensemble
Eq.(\ref{MFmat},\ref{fl},\ref{ring}) and the critical 1D sigma-model
Eq.(\ref{sigma}) turns out to be the simplest {\it generalization} of the
WD
theory that
retains the Tomonaga-
Luttinger liquid analogy extended for finite `temperatures'
$T=1/g^{*}$ which are related with the spectrum of fractal dimensions
Eq.(\ref{weak}).

The Wigner-Dyson two-level statistics for all three symmetry classes can
be expressed through the single kernel $K(s)=\sin(\pi s)/(\pi s)$ in the
following way \cite{RMT}:
\begin{eqnarray}
\label{ALLu}
R^{u}(s)=-K^{2}(s),\\ \label{ALLo}
R^{o}(s)=-K^{2}(s)-\frac{dK(s)}{ds}\,\int_{s}^{\infty}K(x)\,dx,\\
\label{ALLs}
R^{s}(s)=-K^{2}(s)+\frac{dK(s)}{ds}\,\int_{0}^{s}K(x)\,dx. 
\end{eqnarray}
It turns out that such a representation is also valid for the critical
TLCF at $g^{*}\gg 1$ if the kernel is replaced by:
\begin{equation}
\label{kern}
K(s)=\frac{T}{\alpha}\,\frac{\sin(\pi\alpha s)}{\sinh(\pi T
s)},\;\;\;\;T=\frac{1}{g^{*}}. 
\end{equation}
where $\alpha=1$ for the orthogonal and the unitary ensemble and
$\alpha=2$ for the symplectic ensemble. The form of the kernel Eq.(28)
can be guessed from the well known density correlation function for the
case of free fermions in one dimension at a finite temperature that
corresponds to the unitary ensemble \cite{KMut,SLA,MNS}.
In order to prove this statement we note that for $s\gg 1$ and $g^{*}\gg
1$
Eqs.(\ref{ALLu}-\ref{ALLs}) with the kernel Eq.(\ref{kern}) give the same
leading terms as Eqs.(\ref{AAu}-\ref{AAs}) with $G(s,0)$ given by
Eq.(\ref{F}). For $s\ll g^{*}$ Eqs.(\ref{ALLu}-\ref{ALLs}) give 
corrections to the WD statistics that coincide with Eq.(\ref{small}).
Thus the representation Eqs.(\ref{ALLu}-\ref{ALLs}) with the kernel
Eq.(\ref{kern}) are the correct asymptotic expressions for {\it both}
$s\gg 1$ and $s\ll g^{*}$. At $g^{*}\gg 1$ these regions have a
parametrically large overlap so that  Eqs.(\ref{ALLu}-\ref{ALLs})  
are valid for all values of $s$. Given that for all $s$ at $T=0$
and for large energy separations $s\gg 1$ at $T\ll 1$
expressions
Eqs.(\ref{ALLu}-\ref{ALLs})
correspond to equal-time correlations in the Calogero-Sutherland
model, one can expect  Eqs.(\ref{ALLu}-\ref{ALLs}) with the
kernel
Eq.(\ref{kern}) to describe the low-temperature equal-time correlations in
the
Calogero-Sutherland 
model \cite{CSuth} for for all $s$ and the interaction constant
$K=2/\beta$.

One can use the kernel Eq.(28) to compute the spacing distribution
function $P(s,g^{*})$ which can be compared with the corresponding
distribution function $P_{c}(s)$ obtained by the numerical diagonalization
of the three-dimensional Anderson model at the critical point. Such a
comparison is done in Ref.\cite{N} for $\beta=1,2,4$. It turns out that
identifying the parameter $g^{*}$ from the fitting $P(s,g^{*})\sim
e^{-\kappa(g^{*})s}$ with the far exponential tail of $P_{c}(s)\sim
e^{-\kappa s}$ one reproduces the entire distribution function $P_{c}(s)$
extremely well.

 A. M. T. acknowledges a kind hospitality of Abdus Salam 
ICTP where this work was
 performed. 

\widetext
\end{document}